\documentclass[aps,prd,twocolumn,groupedaddress,showpacs]{revtex4}

\usepackage{graphicx} \usepackage{color} \usepackage{natbib}

\usepackage{amsmath,amssymb,amsfonts}
\usepackage{epsf}
\usepackage{epsfig}
\usepackage{longtable}
\begin{document}

%\preprint{}

\title{Testing approximations of thermal effects in neutron star merger simulations}


\author{A.~Bauswein, H.-T.~Janka, R.~Oechslin}
%\thanks{}
\affiliation{Max-Planck-Institut f\"ur
  Astrophysik, Karl-Schwarzschild-Str.~1, D-85748 Garching, Germany}


\date{\today}

\begin{abstract}
We perform three-dimensional relativistic hydrodynamical calculations of neutron star mergers to assess the reliability of an approximate treatment of thermal effects in such simulations by combining an ideal-gas component with zero-temperature, microphysical equations of state. To this end we compare the results of simulations that make this approximation to the outcome of models with a consistent treatment of thermal effects in the equation of state. In particular we focus on the implications for observable consequences of merger events like the gravitational-wave signal. It is found that the characteristic gravitational-wave oscillation frequencies of the postmerger remnant differ by about~50 to 250~Hz (corresponding to frequency shifts of 2 to 8 per cent) depending on the equation of state and the choice of the characteristic index of the ideal-gas component. In addition, the delay time to black hole collapse of the merger remnant as well as the amount of matter remaining outside the black hole after its formation are sensitive to the description of thermal effects.
\end{abstract}

%\keywords{}

\maketitle

\section{Introduction}
High-density matter above nuclear saturation density as it is present in the cores of compact stars usually occurs at temperatures that are low compared to the Fermi energy. Only during the formation process in a stellar core collapse accompanied by a supernova explosion are temperatures of several 10~MeV reached, where, however, because of neutrino losses the temperature drops below 1~MeV within the first minute. Higher temperatures can only be reached again in neutron star binary systems at the end of a tens to hundreds of Myrs lasting inspiral phase, where due to gravitational-wave emission the orbit shrinks and a merging of the two binary components takes place (see~\cite{2009arXiv0912.3529D} for a review).

About ten binary neutron stars are known in our Galaxy, while the actual population is assumed to be much larger due to selection effects \cite{2004Sci...304..547S,2004ApJ...601L.179K,2006astro.ph..8280K}. Although no direct observations of the unavoidable merging processes have been made, it is clear from simulations that the temperature rises up to some~10 to 100~MeV in the merged objects (see e.g.~\cite{2007A&A...467..395O}). It is also known that the additional pressure support due to thermal effects influences the structure and development of the merger remnants \cite{2008PhRvD..78h4033B,PhysRevD.81.024012}.

The modeling of neutron star mergers is limited by the incomplete knowledge of the equation of state of high-density matter. A large variety of EoSs have been proposed based on various theoretical descriptions and matched to various observational data of nuclear matter (see e.g.~\cite{2007ASSL..326.....H}). Neutron stars described by these EoSs show a big spread of the stellar properties, e.g. of the compactness and the maximal mass that can be supported against gravitational collapse \cite{Lattimer:2004pg}. (The astronomical observations of theses properties are not conclusive to constrain the EoSs significantly \cite{2007PhR...442..109L}, but see \cite{2010arXiv1005.0811S}.) Most EoSs consider the ground state of matter, i.e. at zero temperature and in equilibrium with respect to weak interactions (in the following referred to as ``cold'' EoSs). However, in the case of compact star coalescences thermal effects are of significant influence \cite{2007A&A...467..395O,2008PhRvD..78h4033B,PhysRevD.81.024012}. Currently, there are only two microphysical EoSs that include nonzero temperature effects and that are in use for such kind of studies (see~\cite{2009arXiv0912.3529D} for a review): the model by~\cite{1991NuPhA.535..331L}, which we will refer to as LS in the following, and the calculations by~\cite{1998NuPhA.637..435S}, which will be called Shen. In addition, an approximate treatment of thermal effects in combination with any cold microphysical EoSs has been employed e.g. in~\cite{2005PhRvL..94t1101S,2005PhRvD..71h4021S,2006PhRvD..73f4027S,2009PhRvD..80f4037K}, in particular in current fully relativistic studies with microphysical EoSs. However, the reliability of this ideal gas-like extension has not yet been explored.

It is the purpose of this paper to investigate the validity of this approximate description of thermal effects. An understanding of the limitations of this approach is important for the interpretation of observationally relevant results of simulations adopting this scheme. This includes especially the gravitational-wave signals of merger events, which may be detectable with current and upcoming gravitational-wave detectors like LIGO \cite{Abbott:2007kv2} and VIRGO \cite{Acernese:2006bj}. Such experiments have some potential to constrain the high-density EoS once measurements become available. These attempts will have to rely on the results of simulations with various theoretical prescriptions of the EoS using partially the approximative temperature implementation (e.g.~\cite{2005PhRvL..94t1101S,2005PhRvD..71h4021S,2006PhRvD..73f4027S,2009PhRvD..80f4037K}), which makes it important to estimate the influence of this simplified description. Furthermore, neutron star mergers are speculated to be the progenitor systems of short gamma-ray bursts, and the modeling of the conditions for producing such gamma-ray bursts may also depend on the particular treatment of thermal effects \cite{2006MNRAS.368.1489O,2008PhRvD..78h4033B,PhysRevD.81.024012}.

First we briefly review the relevant earlier findings, which our analysis is supposed to supplement. Our methods are introduced in Sect.~\ref{sec:meth} and the properties of the EoSs used in this study are discussed in Sect.~\ref{sec:eos}. Sect.~\ref{sec:sim} presents the temperature phenomenology in merger simulations. In Sect.~\ref{sec:ana} we investigate the influence of the approximate treatment of thermal effects on observational features of the merging process, and Sect.~\ref{sec:con} summarizes our conclusions.

\section{Review of previous studies} \label{sec:rev}

Different approximations of microphysical EoSs have been compared in previous studies, which we briefly summarize here to set the stage for our analysis. We focus on those works that examine and demonstrate the importance of including thermal effects in neutron star merger simulations.

The simplest approach to describe neutron star matter is the use of the polytropic relation $P=\kappa\rho^{\Gamma}$, which determines the pressure $P$ as dependent on the rest-mass density $\rho$ with constant $\kappa$ and $\Gamma$. Employing this barotropic function during the evolution of a neutron star merger corresponds to the extreme case of perfectly efficient cooling, since this EoS does not allow for shock heating (``isentropic case''). This means that thermal energy is taken away from the matter instantaneously, and in fact the energy equation does not need to be evolved. As an alternative, one can use an ``ideal-gas'' ansatz, where the pressure is given by $P=(\Gamma-1)\epsilon\rho$ with the internal specific energy $\epsilon$. In this case the energy is conserved also when shocks develop. This approach is particularly interesting if the initial data are constructed also by a polytropic relation and a comparison between isentropic and ideal gas simulations is made. This allows one to judge the importance of thermal effects. Both EoS treatments were compared in \cite{2008PhRvD..78h4033B}, where the lower pressure support in the isentropic case led to an earlier gravitational collapse of the merger remnant compared to the ideal-gas calculation. Furthermore, the matter remaining outside the black hole formed a thin disk-like structure, while the ideal-fluid EoS resulted in a vertically inflated torus. This finding can also be explained by the additional pressure support provided by the nonzero temperature, which ``inflates'' the low-density material in the outskirts of the remnant.

In Ref.~\cite{2007A&A...467..395O} it was investigated to which extent an analytical ideal-gas EoS can be used to reproduce the features of a neutron star merger simulation with a microphysical EoS. The initial stars were set up as polytropes, with $\kappa$ and $\Gamma$ fixed in a way that the relation between central density and gravitational mass of isolated neutron stars matched the relation of the microphysical EoS in a wide range. It was found that the simplified treatment yields a remnant structure and a density profile similar to the one obtained with the microphysical EoS. The gravitational-wave emission however, was only in qualitative agreement \cite{2007PhRvL..99l1102O}, and not all relevant features were compatible between the simulations. Moreover, these conclusions were based on only one binary configuration, and it remains unanswered whether the similarities also hold for a larger variety of models. Furthermore, the study was performed for only one microphysical EoS and it should be explored whether this approximation works also for other microphysical EoSs.

In \cite{2007A&A...467..395O,PhysRevD.81.024012} it was analyzed whether the inclusion of thermal effects is at all  important for the remnant properties and observable quantities in neutron star merger simulations when microphysical EoSs are used. The comparisons presented there are closely related to the aforementioned in \cite{2008PhRvD..78h4033B}. The data of a temperature-dependent microphysical EoS were implemented in a way to impose perfectly efficient cooling, i.e. only the zero-temperature sector of the EoS was taken into account. This model was compared with simulations employing the same microphysical EoS, but using its full temperature dependence. As in \cite{2008PhRvD..78h4033B} the major differences are explained by the lower thermal pressure contribution in the zero-temperature calculations. While in the case of the temperature-dependent simulation an inflated, ``hot'' torus was obtained, the omission of thermal effects led to a disk with a small vertical extension and a higher density. In addition, the ``cooling losses'' in the zero-temperature models resulted in more compact merger remnants, which shifted the gravitational-wave frequencies in the postmerger stage to higher values. Moreover, it also increased the amount of matter remaining outside the black hole after the gravitational collapse of the central object~\cite{2007PhRvL..99l1102O,PhysRevD.81.024012}.

In summary, the studies mentioned here show that the consideration of nonzero temperature effects is generally important in simulations of neutron star mergers. Therefore, in the following we will analyze a more sophisticated approach to include thermal effects in an approximate manner in simulations with microphysical EoSs. It was originally introduced in \cite{1993A&A...268..360J} and has been applied in the context of stellar core-collapse (e.g.~\cite{2005PhRvD..71f4023D}) and in neutron star merger studies (e.g.~\cite{2005PhRvL..94t1101S,2005PhRvD..71h4021S,2006PhRvD..73f4027S,2009PhRvD..80f4037K}). The ansatz provides an additional pressure contribution $P_{\mathrm{th}}=(\Gamma_{\mathrm{th}}-1)e_{\mathrm{th}}$ due to thermal effects, which is proportional to the thermal energy density $e_{\mathrm{th}}$ assuming an ideal-gas like behavior with a constant ``ideal-gas index'' $\Gamma_{\mathrm{th}}$. The value of $\Gamma_{\mathrm{th}}$ cannot be unambiguously determined from physics arguments but was chosen more or less ad hoc. It will be argued in Sect.~\ref{sec:eos} (see Fig.~\ref{fig:gammath}) that assuming a constant ideal-gas index the applicability of the ansatz is at least questionable and requires an investigation. Furthermore, we will explain in Sect.~\ref{sec:meth} that for barotropic EoSs this approximate description implies an unphysical variation of the electron number, which affects spuriously the EoS properties and thus may have an impact on the dynamics of a neutron star merger.

\section{Methods and setup} \label{sec:meth}
We perform simulations of neutron star mergers using a general relativistic Smoothed Particle Hydrodynamics scheme, which employs the conformal flatness approximation of general relativity \cite{1980grg..conf...23I}. The description of the code can be found in \cite{2002PhRvD..65j3005O,2007A&A...467..395O}. Various microphysical EoSs are implemented in our model to close the set of hydrodynamical equations. The EoS provides the pressure $P$ and the specific internal energy density $\epsilon$ as a function of the rest-mass density $\rho$, the temperature $T$ and the electron fraction $Y_{\mathrm{e}}$. The rest-mass density, the energy density and the electron fraction are given by the hydrodynamical evolution. We assume the advection of the initial electron fraction and neglect neutrino production, which leads to the simple evolution equation $\frac{dY_{\mathrm{e}}}{dt}=0$. A first look-up in the EoS determines the temperature as a function of the energy density for the given $\rho$ and $Y_{\mathrm{e}}$ by a numerical inversion procedure. Then the pressure is found from the EoS as dependent on $\rho$, $Y_{\mathrm{e}}$ and $T$. Both operations involve numerical interpolation schemes, because the EoS is given in the form of a table.

For this study we employ the Shen EoS and the LS EoS. These EoSs provide the full temperature dependence and
therefore, they can be used for a comparison to quantify the
deviations between a consistent and an approximate inclusion of
thermal effects.

The approximative description of thermal effects can be applied for any
cold, microphysical EoS. For this reason it was included in simulations of neutron
star mergers. The zero-temperature EoSs provide an unambiguous relation between the rest-mass density and the pressure, as well as the rest-mass density and
the energy density. Typically, these functional dependences are given in the form of a table. For barotropic EoSs that are commonly used, the electron fraction is fixed by equilibrium with respect to weak interactions, i.e. the evolution of the advection equation $\frac{dY_{\mathrm{e}}}{dt}=0$ used in consistent models is abandoned \footnote{Note that this is in contrast to the zero-temperature approach taken in \cite{PhysRevD.81.024012}, where the full $T=0$ slice of the EoS, i.e. $P(\rho,Y_{\mathrm{e}},T=0)$ and $\epsilon(\rho,Y_{\mathrm{e}},T=0)$, was used, and $Y_{\mathrm{e}}$ of the mass elements was advected with time and the advection of $Y_{\mathrm{e}}$ was followed in time.}.

In order to include
temperature effects one splits up the pressure and the specific
internal energy into a cold and a thermal part:
\begin{eqnarray}
P&=&P_{\mathrm{cold}}+P_{\mathrm{th}},\label{eq:Psplit}\\
\epsilon&=&\epsilon_{\mathrm{cold}}+\epsilon_{\mathrm{th}}.\label{eq:esplit}
\end{eqnarray}
The cold contributions are taken from the tabulated, microphysical, zero-temperature
EoS and are functions of the rest-mass density $\rho$ only. The rest-mass density $\rho$ and the specific internal
energy $\epsilon$ are given by the hydrodynamical evolution. Then, for a
given $\epsilon$ and $\rho$ one defines the thermal part of the
specific internal energy by
\begin{equation}
\epsilon_{\mathrm{th}}=\epsilon-\epsilon_{\mathrm{cold}}(\rho),
\end{equation}
where the cold contribution is taken from the cold microphysical EoS table determined by $\rho$.
Assuming an ideal-gas like behavior the thermal contribution to the
pressure is set to
\begin{equation}
P_{\mathrm{th}}=(\Gamma_{\mathrm{th}}-1)\rho\epsilon_{\mathrm{th}},\label{eq:pth}
\end{equation}
with the ideal-gas index $\Gamma_{\mathrm{th}}$, which is
assumed to be constant for all $\rho$ and $\epsilon$. Now one
obtains the pressure from Eq.~\eqref{eq:Psplit}, which closes the
hydrodynamical equations. By means of this scheme arbitrary barotropic
EoSs can be implemented in hydrodynamical simulations consistently with
the first law of thermodynamics.

Besides the approximate treatment of temperature effects, a further, less obvious, approximation is implied by this implementation, when used with barotropic EoSs. The cold EoSs are constructed for matter in neutrino-less beta-equilibrium, i.e. for a neutrino chemical potential of zero. This means that the pressure and the energy density are functions of $\rho$ only, and the explicit dependence of both on $Y_{\mathrm{e}}$ is neglected. Therefore, when matter is compressed or expanded, $Y_{\mathrm{e}}$ changes unphysically due to the imposed condition of beta-equilibrium (because for a barotropic EoS $Y_{\mathrm{e}}$ in neutrino-less beta-equilibrium effectively is a function of $\rho$ only). However, in reality, matter that was initially in beta-equilibrium, would be out of equilibrium after compression or decompression, because the composition changes on a longer than the dynamical timescale \footnote{More precisely, when matter heats up by compression or shocks during the merger, it becomes intransparent to neutrinos. Therefore, beta-equilibrium with a nonzero population of neutrinos develops quickly on weak interaction timescales, but the evolution to neutrino-less beta-equilibrium occurs only on the much longer diffusion and cooling timescale. See \cite{1966ApJ...145..514M} for a discussion of the timescales to reestablish neutrino-less beta-equilibrium in the context of neutron-star oscillations.}. Hence, the instantaneously enforced adjustment to beta-equilibrium leads to spurious effects in the dependent quantities of the EoS like the pressure and the energy density.

For the sake of a comparison we construct cold EoSs from the tables of the LS and the Shen models, i.e. we keep only the information about the microphysical properties at $T=0$ and beta-equilibrium. In fact, we use the barotropic relations that are also employed for computing the initial data of our fully consistent simulations.

For both EoSs (LS and Shen) we perform three runs each: one using the full temperature dependence of the EoS, and two simulations where the approximate treatment of thermal effects is employed in combination with the cold EoS (see Tab.~\ref{tab:models}). One calculation is done with $\Gamma_{\mathrm{th}}=1.5$, the other with $\Gamma_{\mathrm{th}}=2$. The choices for $\Gamma_{\mathrm{th}}$ will be discussed in Sect.~\ref{sec:eos}.

For every model we simulate the merging of two non-spinning stars with gravitational masses in isolation of 1.35~$M_{\odot}$. This setup can be considered as a canonical configuration that is predicted to be very abundant in the neutron star binary population~\cite{2008ApJ...680L.129B}. Furthermore, for both EoSs this set of binary parameters results in the formation of a so-called hypermassive postmerger remnant, a differentially rotating object that collapses to a black hole after angular momentum redistribution on typical timescales of some 10 milliseconds. Higher total binary masses would cause the prompt formation of a black hole shortly after the stars have come into contact. Since temperature effects are important in particular during the formation and the evolution of a hypermassive postmerger remnant, we focus on the aforementioned equal-mass binaries with moderate neutron star masses. Our analysis is constrained to this limited, but representative, set of models because of the high demand of computational resources. In particular, we found it better to compute until black hole formation if possible for the chosen configurations, rather than conducting more models for a shorter time. Therefore, our investigation should be considered as a matter-of-principle study rather than a detailed and full survey of uncertainties in the parameter space (masses, spins, EoS) of binary setups.

Two additional simulations with different resolutions are performed for the LS EoS using the consistent temperature description to assess the resolution dependence of the quantities we are interested in. For the characteristics of a merger discussed below, Tab.~\ref{tab:models} gives the results of runs with about 140,000 and 340,000 instead of our standard resolution of about 550,000 SPH particles. They show a strong dependence of the estimated torus mass on the resolution (deviations to the run with the standard resolution of about 97 per cent for the lowest resolution and of about 14 per cent for the intermediate resolution). The delay time to the black hole formation depends moderately on the resolution, while the oscillation frequency of the postmerger remnant is less sensitive to the particle number. However, we stress that for each set of models that will be discussed in the following, based either on the Shen EoS or the LS EoS we use exactly the same initial data and the standard resolution. Therefore, deviations among these calculations are entirely caused by the different treatments of thermal effects.
 \begin{table}%[H] add [H] placement to break table across pages
 \caption{\label{tab:models}Models discussed in the text. The oscillation frequency of the postmerger remnant is
   denoted by $f_{\mathrm{peak}}$.  $\tau_{\mathrm{delay}}$ is the
   delay time between the merging and the formation of a black hole. Note that
   in the cases where the black hole collapse does not occur during the
   simulation, we give a lower limit for $\tau_{\mathrm{delay}}$. This
   value is determined by the finite simulation time and may differ
   significantly from the true value (see \cite{rezzolla}). The estimated mass of the torus remaining outside the black hole after the gravitational collapse is listed in the last column. For the simulations based on the Shen EoS we give the amount of matter fulfilling the torus criterion 16~ms after the merging, because for these models a black hole does not form within the simulation time. ``LS 140K'' and ``LS 340K'' refer to runs with lower numerical resolutions using the full LS EoS table. To estimate resolution effects these simulations should be compared to the entries in the first row.}
 \begin{ruledtabular}
 \begin{tabular}{|l|l|l|l|}
EoS& $f_{\mathrm{peak}}$~[kHz]&$\tau_{\mathrm{delay}}$~[ms]&$M_{\mathrm{torus}}~[M_{\odot}]$ \\
\hline \hline
LS full table& 3.24& 20.8& 0.064\\ \hline
LS $\Gamma_{\mathrm{th}}=1.5$&3.38& 10.3& 0.045\\ \hline
LS $\Gamma_{\mathrm{th}}=2$&2.99& 21.0& 0.054\\ \hline
Shen full table& 2.21& $>$18.1& 0.075\\ \hline
Shen $\Gamma_{\mathrm{th}}=1.5$& 2.27& $>$16.8& 0.084\\ \hline
Shen $\Gamma_{\mathrm{th}}=2$&2.16& $>$18.0& 0.074\\ \hline
LS 140K& 3.22& 17.5& 0.126\\ \hline
LS 340K& 3.25& 18.4& 0.073\\
 \end{tabular}
 \end{ruledtabular}
\end{table}

\section{EoS properties} \label{sec:eos}

\begin{figure}
\includegraphics[width=8.9cm]{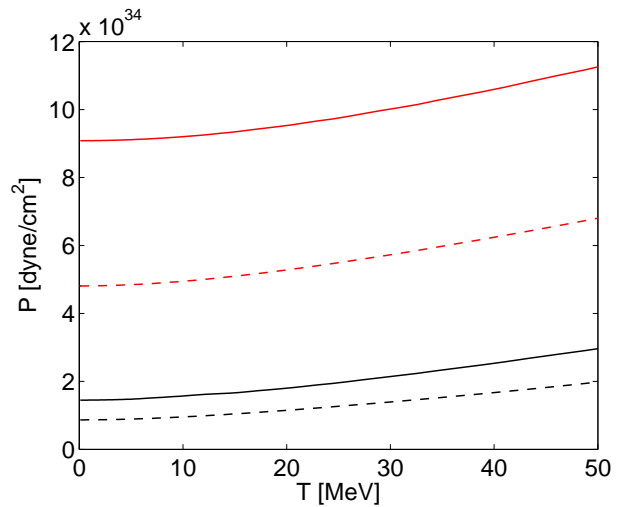}
\caption{\label{fig:PT}Pressure as a function of the temperature for
  rest-mass densities of $\rho=3\cdot 10^{14}\mathrm{g/cm^3}$ (black
  curves) and $\rho=6\cdot 10^{14}\mathrm{g/cm^3}$ (red curves). The pressure is shown for a constant value of the electron fraction as determined from the condition of beta-equilibrium at $T=0$ for the given densities. The solid lines
  correspond to the Shen EoS, while the dashed curves display the pressure for
the LS EoS.}
\end{figure}

The dependence of the pressure on the temperature is shown in Fig.~\ref{fig:PT} for both EoS models used in this study. Two constant values of the rest-mass density, $\rho=3\cdot 10^{14}\mathrm{g/cm^3}$ (black curves) and $\rho=6\cdot 10^{14}\mathrm{g/cm^3}$ (red curves), were chosen corresponding to about one and two times nuclear saturation density. For low temperatures (below 5~MeV) the pressure is nearly constant as a consequence of the degeneracy of nuclear matter at these densities. One also recognizes that the Shen EoS provides a higher pressure at the same density. For the dependence of the pressure on the rest-mass density and an overview of the mass-radius relations of neutron stars described by the soft LS EoS and the stiff Shen EoS we refer the reader to Figs.~1 and~2 in~\cite{PhysRevD.81.024012}. As a consequence of these EoS properties the stars described by the Shen model are in general less compact. In addition, \cite{PhysRevD.81.024012} provides information about the influence of thermal effects on the stellar structure: for temperatures of some 10 MeV the stellar radii are a few kilometers bigger.

\begin{figure}
\begin{center}
\includegraphics[width=8.9cm]{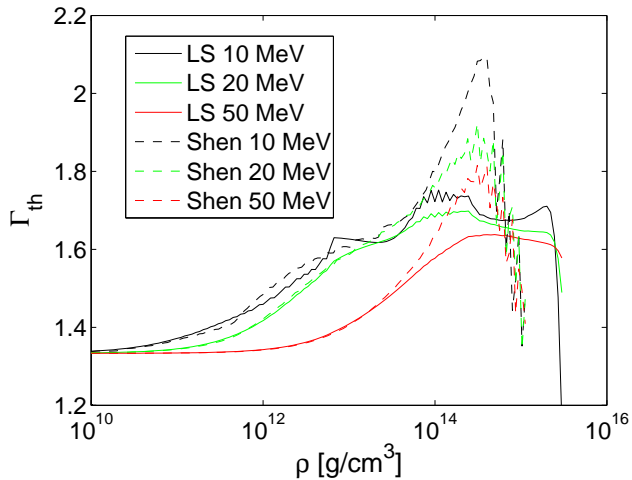}
\caption{\label{fig:gammath} ``Ideal-gas index'' $\Gamma_{\mathrm{th}}$ as a function of the rest-mass density for the Shen EoS and the LS EoS for typical temperatures. See Eq.~\ref{eq:gammath} for the definition of $\Gamma_{\mathrm{th}}$. The electron fraction $Y_{\mathrm{e}}$ for a given density is determined from the condition of neutrino-less beta-equilibrium at $T=0$.}
\end{center}
\end{figure}

For a given $\rho$, $T$ and $Y_{\mathrm{e}}$ one can compute the local effective ``ideal-gas index'' $\Gamma_{\mathrm{th}}$ of the EoSs. According to Eqs.~\eqref{eq:Psplit}, ~\eqref{eq:esplit} and ~\eqref{eq:pth} one finds
\begin{equation} \label{eq:gammath}
\Gamma_{\mathrm{th}}(\rho,T,Y_{\mathrm{e}})=\frac{P_{\mathrm{th}}}{\rho\epsilon_{\mathrm{th}}}+1=\frac{P(\rho,T,Y_{\mathrm{e}})-P(\rho,0,Y_{\mathrm{e}})}{\rho\left[\epsilon(\rho,T,Y_{\mathrm{e}})-\epsilon(\rho,0,Y_{\mathrm{e}}) \right]}+1.
\end{equation}
This thus defined ideal-gas index $\Gamma_{\mathrm{th}}$ is shown as a function of the rest-mass density in Fig.~\ref{fig:gammath} for typical temperatures occurring during the merging of neutron stars (see Sect.~\ref{sec:sim}). The variation of the displayed values of the ideal-gas index reveals that the choice of a constant value of $\Gamma_{\mathrm{th}}$ can only be a very crude approximation and requires a validation, which we aim at with this study. Since we would like to assess the validity of this ansatz in neutron star merger simulations, we adopt the values for $\Gamma_{\mathrm{th}}$ previously used in the corresponding literature~\cite{1993A&A...268..360J,2005PhRvL..94t1101S,2005PhRvD..71h4021S,2006PhRvD..73f4027S,2009PhRvD..80f4037K}, i.e. $\Gamma_{\mathrm{th}}=1.5$ and $\Gamma_{\mathrm{th}}=2$. One should keep in mind that the dynamics of the merger are determined mainly by the supranuclear EoS. Hence, the ideal-gas index should primarily reproduce the thermal EoS behavior in this density range. From Fig.~\ref{fig:gammath} one might conclude that the choice $\Gamma_{\mathrm{th}}=2$ is slightly too high, thus it could overestimate the thermal pressure contribution $P_{\mathrm{th}}$, while $\Gamma_{\mathrm{th}}=1.5$ might yield a too low pressure support. One also recognizes that the ideal-gas index is lower the higher the temperature is. In comparison to the LS EoS, the Shen EoS has higher values of $\Gamma_{\mathrm{th}}$.

Furthermore, for densities below~$\sim 10^{11}\mathrm{g/cm^3}$ the ideal-gas index $\Gamma_{\mathrm{th}}$ approaches 4/3. This is a consequence of the fact that in this density regime the pressure is provided mainly by an ideal gas of ultra-relativistic electrons and photons.

\section{Simulations} \label{sec:sim}
\begin{figure*}
\includegraphics[width=8.5cm]{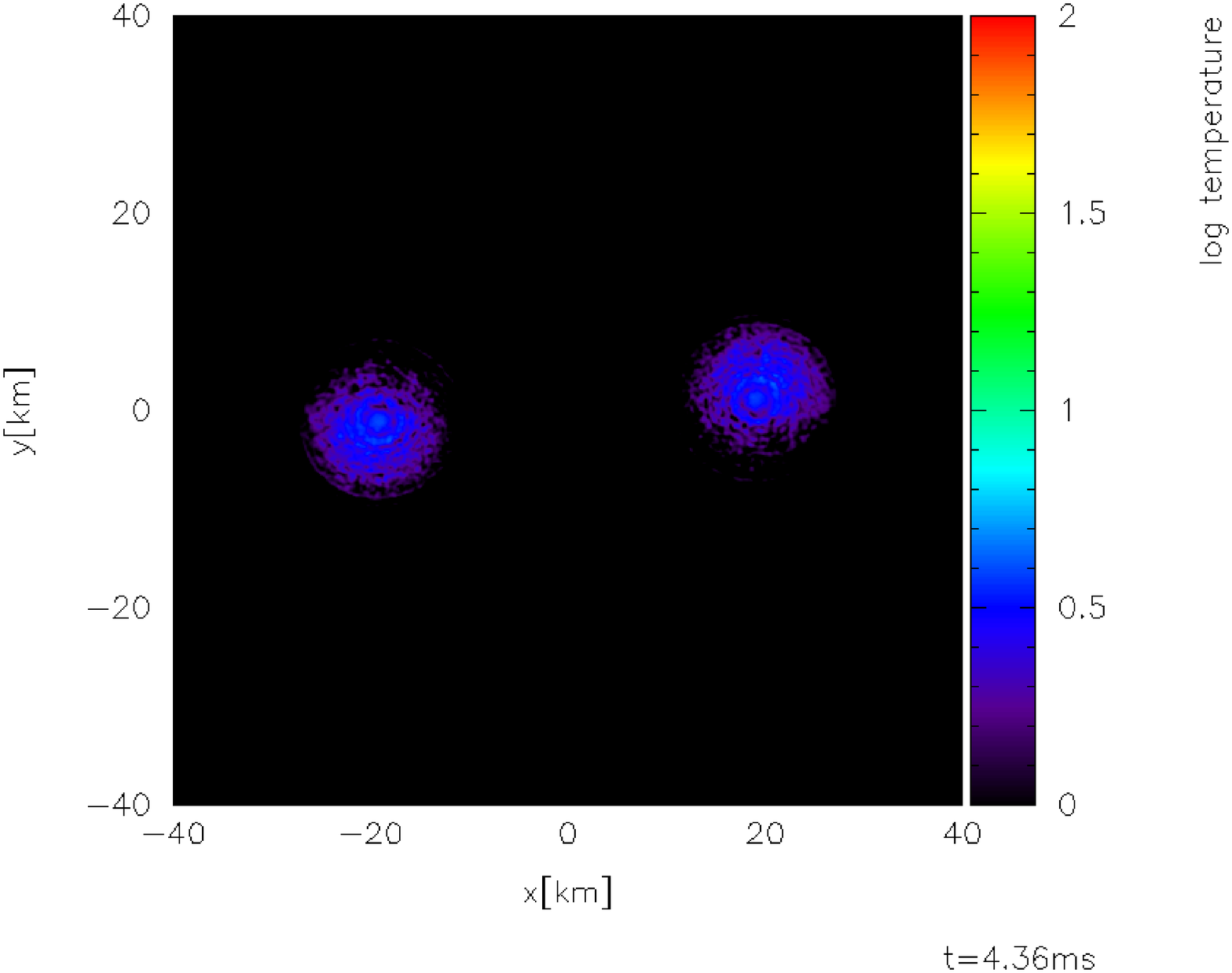}
\includegraphics[width=8.5cm]{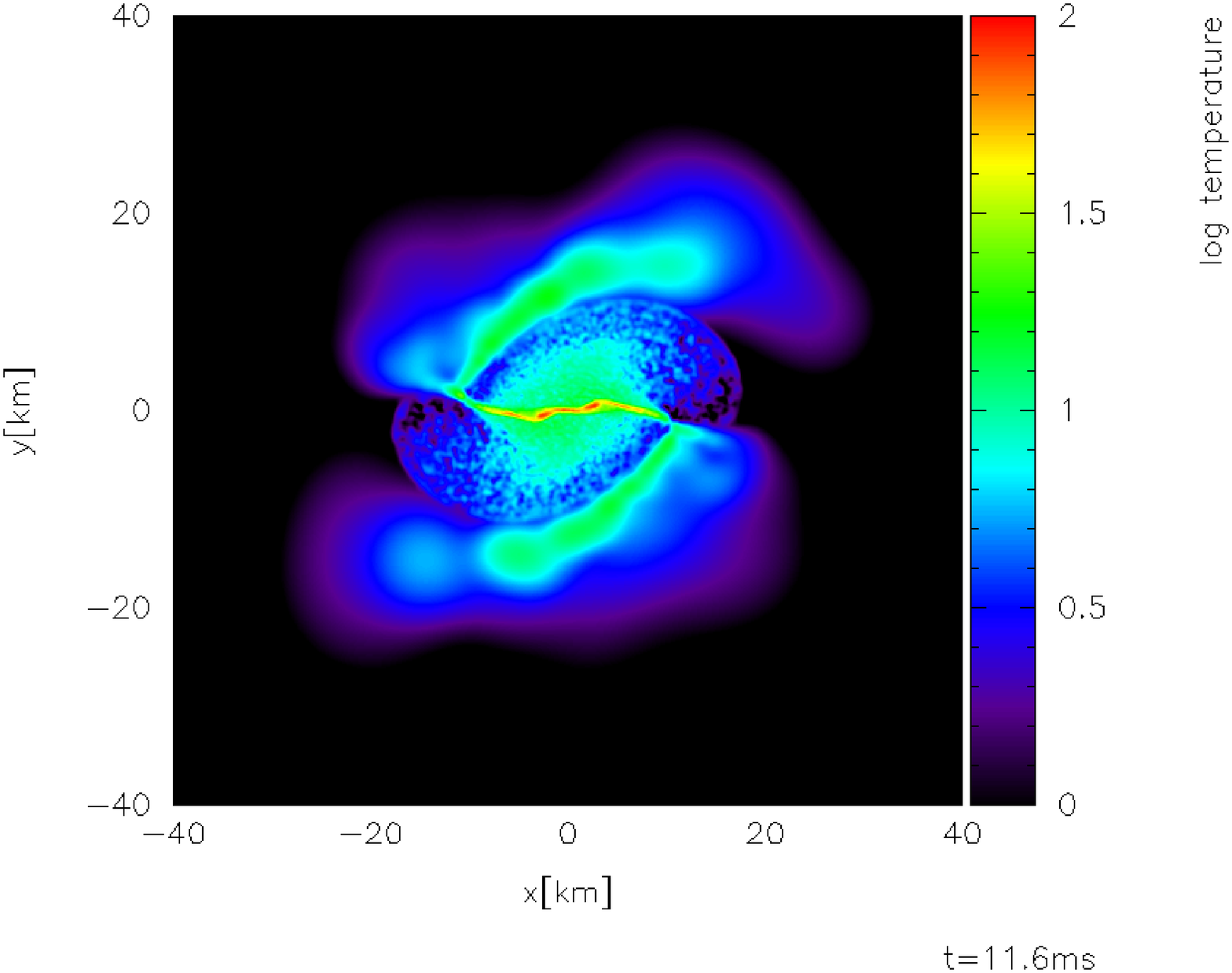}\\
\includegraphics[width=8.5cm]{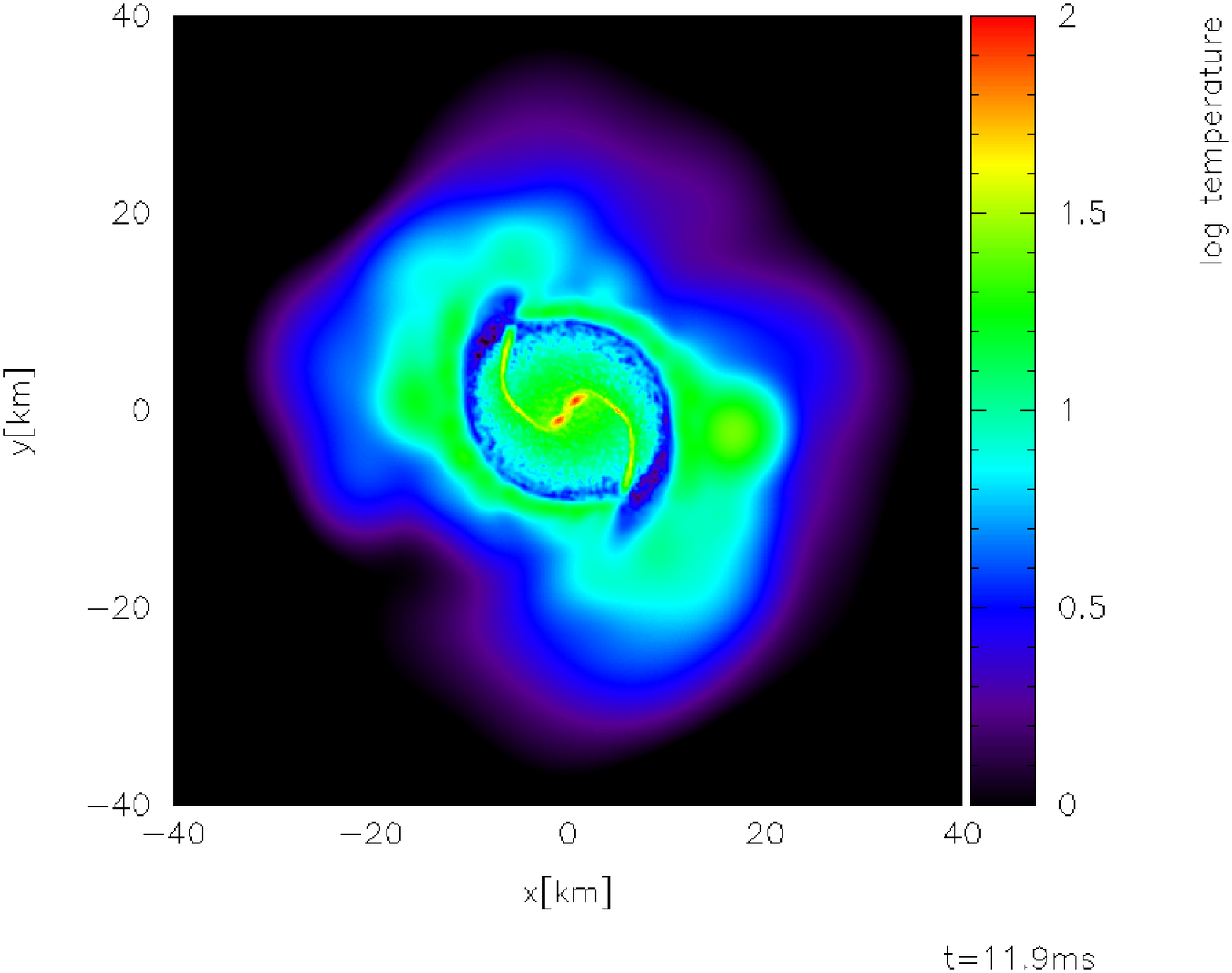}
\includegraphics[width=8.5cm]{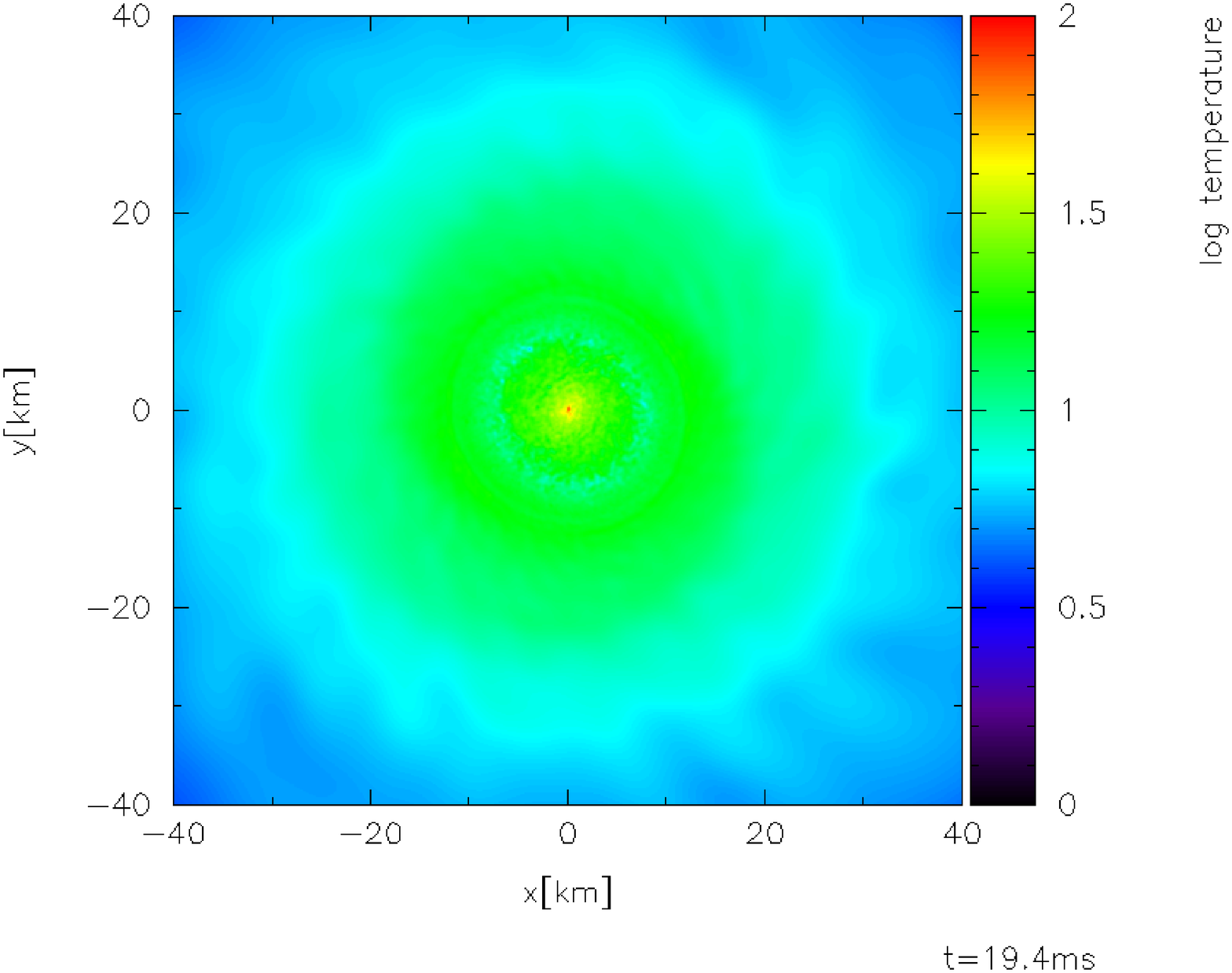}
\caption{
\label{fig:snaptemp}Temperature in MeV in the orbital plane of a merging NS binary with 1.3 $M_{\odot}$ and 1.35 $M_{\odot}$ components  for the LS EoS using the fully consistent temperature implementation. The rotation is counter-clockwise. Note that the temperature is given on a logarithmic scale in order to resolve all relevant features. The plots were created with the visualization tool SPLASH \cite{2007PASA...24..159P}.}
\end{figure*}
We evolve neutron star binaries with our hydrodynamical code through the last phase of the inspiral, the merging and the postmerger stage. These different phases lead to distinctive temperature distributions in the orbital plane of the merger with the LS EoS using the full temperature description (Fig.~\ref{fig:snaptemp}). When the stars orbit around each other, numerical dissipation heats the stellar interior to some MeV (upper left panel). As can be seen in Fig.~\ref{fig:PT}, this temperature increase can be expected to be dynamically unimportant. Then, during the merging, shock heating produces temperatures up to about 100~MeV at the contact layer between the stars, where also Kelvin-Helmholtz shear vortices develop (upper right). The subsequent compression leads to an average temperature of about 20~MeV in the dense core of the central object (upper right and lower left panels). Moreover, material is shed off from the merger remnant to form a dilute halo around the hypermassive object. The halo matter is shock heated by the interaction with the surface of the deformed, dense, rotating remnant, leading to temperatures of 20 to 30~MeV (upper right and lower left panels). Because of the differential rotation the shock-heated, hot material from the clash of the two stars is spread within the central object (lower left panel). After some milliseconds an approximately stationary configuration is reached with a nearly axisymmetric temperature distribution (lower right panel). Remarkably, the outer part of the central core remains at a slightly lower temperature than the inner core and the surrounding medium (barely visible), because it is not as strongly compressed and it is not subject to shock heating. During the whole evolution after the merging the temperature of the supranuclear matter forming the central object is above 20~MeV reaching even more than 60~MeV at the center. From this analysis it is clear that the effects of nonzero temperature should be considered as important for the dynamics and the shape of the postmerger structures (see Fig.~\ref{fig:PT} and the summary in Sect.~\ref{sec:rev}).

Basically the same picture arises for the simulation with the Shen EoS, where the merging proceeds similarly to the LS case. However, the temperatures are lower. When the remnant has settled to an axisymmetric configuration the central temperature is about 30~MeV, while the outer remnant parts have temperatures between 10 and 20~MeV. This can be understood from the fact that the stiffer EoS leads to a less compact remnant structure, thus a less extreme compression of the matter.

For the simulations using the approximate description of thermal effects, a relation between the thermal energy and the temperature is not provided by the EoS and therefore temperature information cannot be given on this level. Only by introducing an additional assumption about the nature of the hot matter, for instance considering it as a gas of nucleons, one can estimate the temperature (see e.g.~\cite{2005PhRvD..71h4021S}). We refrain from doing this here.

\section{Results}\label{sec:ana}
In this section we discuss the consequences of an approximative ideal-gas description of thermal EoS contributions for observable features of binary neutron star mergers.
\subsection{Gravitational waves}
The overall dynamics of the $\Gamma_{\mathrm{th}}$-runs are similar to
the corresponding simulations using the full temperature
dependence. However, in order to judge the quality of the
approximate treatment of thermal effects, one should investigate the
implications of the different approaches for observable
aspects. Therefore, we focus in our analysis on the
gravitational-wave signal, which is the most direct probe of neutron
star collision events.

\begin{figure}
\begin{center}
\includegraphics[width=8.9cm]{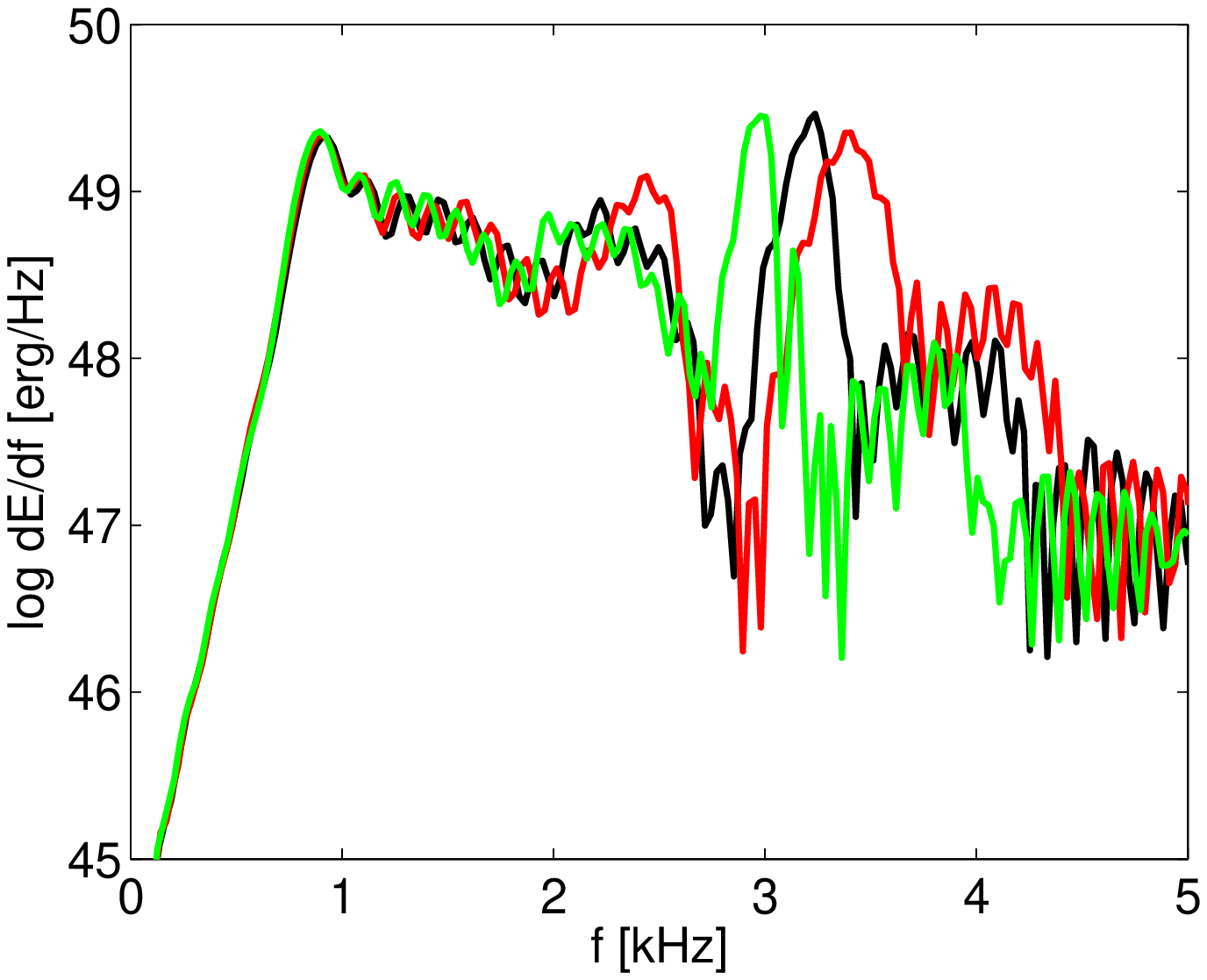}
\caption{\label{fig:FFT-LS} Direction and polarization averaged gravitational-wave luminosity spectra for symmetric binaries with $M_1=M_2=1.35~M_{\odot}$ described by the LS EoS using the full temperature dependence (black curve) and an approximate description of thermal effects with $\Gamma_{\mathrm{th}}=1.5$ (red curve) and $\Gamma_{\mathrm{th}}=2$ (green curve). Note that our simulations start only a few orbits before the merging of the stars. Hence, power from the preceding inspiral phase at lower frequencies ($< 1$~kHz) is missing in this plot.}
\end{center}
\end{figure}

\begin{figure}
\begin{center}
\includegraphics[width=8.9cm]{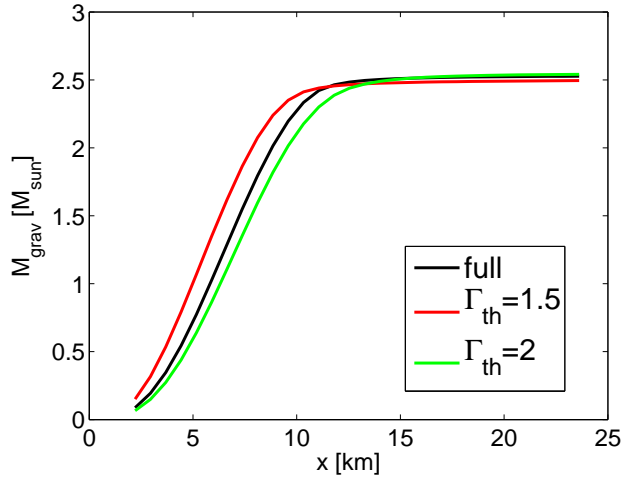}
\caption{\label{fig:encgth}Enclosed gravitational mass within an
  ellipsoid with the semiaxes $a=x$, $b=x$ and $c=x/2$ for the
  1.35~$M_{\odot}$+1.35~$M_{\odot}$ binaries about 8~ms after the merging. Shown are results with the LS EoS using the full temperature dependence (black curve) and an approximate description of thermal effects with $\Gamma_{\mathrm{th}}=1.5$ (red curve) and $\Gamma_{\mathrm{th}}=2$ (green curve). Note that within general relativity the gravitational mass is only defined in isolation. Here $M_{\mathrm{grav}}$ denotes the enclosed contribution to the ADM mass, neglecting the extrinsic curvature terms. Using the enclosed rest mass yields a similar result. The distance $x$ is given in isotropic coordinates.}
\end{center}
\end{figure}

\begin{figure}
\begin{center}
\includegraphics[width=8.9cm]{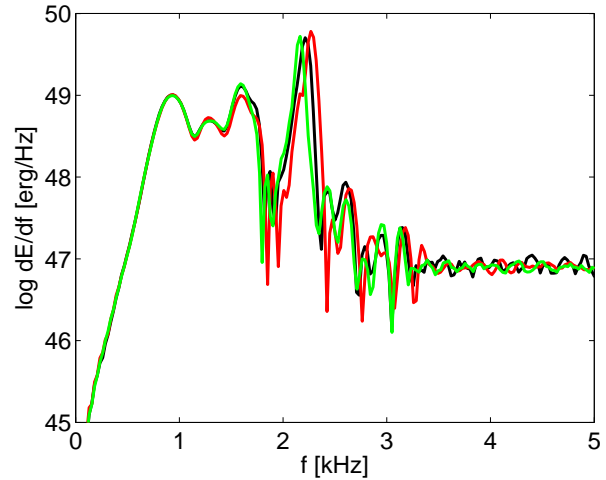}
\caption{\label{fig:FFT-Shen} Same as Fig.~\ref{fig:FFT-LS}, but for the Shen EoS.}
\end{center}
\end{figure}

As can be seen in Fig.~\ref{fig:snaptemp}, temperature effects can only be
dynamically important in the merging and postmerging stages, while they are irrelevant during the inspiral of the binary components. It is known that the gravitational-wave emission of the deformed, oscillating,
hypermassive remnant depends sensitively on EoS properties~\cite{2007PhRvL..99l1102O,PhysRevD.81.024012}. And it was also shown previously that the inclusion of thermal effects has an impact on the gravitational-wave signal of the postmerger phase~\cite{2007PhRvL..99l1102O,2008PhRvD..78h4033B,PhysRevD.81.024012}.

The emission of gravitational radiation from the postmerger remnant is
dominated by a narrow frequency band around a frequency $f_{\mathrm{peak}}$, which is visible as a prominent peak structure in the spectrum (see Fig.~\ref{fig:FFT-LS}). Hence, this peak frequency $f_{\mathrm{peak}}$ can be used as a quantitative feature of the gravitational-wave signal in order to characterize the emission during the postmerger phase.

In Fig.~\ref{fig:FFT-LS} one can read off the peak frequencies of the LS models from the gravitational-wave luminosity spectrum, which is given by $dE_{GW}/df=2\pi ^2 D^2 f^2 \langle|\tilde{h}_+|^2+|\tilde{h}_\times|^2\rangle$. Here $D$ is the distance from the source, which cancels out when multiplied with the Fourier transformed waveforms $\tilde{h}$ for both polarizations averaged over all emission directions. The amplitudes of the gravitational-wave train are computed by means of a quadrupole formula that takes into account post-Newtonian effects \cite{1990MNRAS.242..289B,2007A&A...467..395O}.

With the LS EoS using the full temperature dependence, the peak frequency of the postmerger oscillation is located at 3.22~kHz (black curve in Fig.~\ref{fig:FFT-LS}). In comparison, the peak frequencies of the simulations employing the approximate scheme are about 200~Hz lower for the $\Gamma_{\mathrm{th}}=2$ model and about 200~Hz higher for the $\Gamma_{\mathrm{th}}=1.5$ run (see Tab.~\ref{tab:models}). To have a reference for judging the importance of such differences, we note that for instance for a merger of a binary with 1.2~$M_{\odot}$ and 1.35~$M_{\odot}$ neutron stars described by the full temperature-dependent LS EoS the peak frequency is 3.04~kHz (see~\cite{PhysRevD.81.024012}). This value differs also by about 200~Hz from our symmetric standard case model.

The discrepancies between the models with the different treatments of thermal effects can be understood as a consequence of the different remnant structures. More precisely, the resulting compactness of the remnants is the crucial property determining $f_{\mathrm{peak}}$. The more compact the objects are, the higher frequencies are obtained. In this context we refer the reader also to~\cite{PhysRevD.81.024012}, where the dependence on the compactness was discussed. Since the $\Gamma_{\mathrm{th}}=1.5$ implementation of thermal effects leads to a lower total pressure support, the remnant is more compact compared to the case with the full EoS. In contrast, $\Gamma_{\mathrm{th}}=2$ overestimates the thermal pressure contribution and consequently, the central object is less compact.

For isolated stellar objects the compactness is defined as
$C=GM/\left( c^2R \right)$, with the gravitational mass $M$ and circumferential radius
$R$ of the star. $G$ and $c$ are the gravitational constant and the
speed of light. For neutron star merger remnants, however, this definition of
the compactness is not useful, because the remnants do not have a
well-defined surface (see Fig.~\ref{fig:snaptemp}). Instead we introduce a measure for the
compactness by considering the enclosed mass within ellipsoids with
the semiaxes  $a=x$, $b=x$ and $c=x/2$. (Note that the hypermassive
object is highly deformed. The defined ellipsoids describe
approximately isodensity surfaces.) Fig.~\ref{fig:encgth} clearly
illustrates the different remnant structures of the three models based
on the LS EoS and confirms our explanation given above. The enclosed
mass is computed about 8~ms after the merging, where the merger time
is defined as the moment when the gravitational-wave amplitude becomes maximal.

Basically the same picture as for the LS EoS holds for the simulations using the Shen EoS. The luminosity spectra are displayed in Fig.~\ref{fig:FFT-Shen}. The generally lower values of $f_{\mathrm{peak}}$ are explained by the fact that the Shen EoS leads to less compact objects than the LS EoS. The differences in the peak frequencies are only about 50 Hz (see Tab.~\ref{tab:models}). Again, assuming $\Gamma_{\mathrm{th}}=1.5$ results in a higher oscillation frequency, while using $\Gamma_{\mathrm{th}}=2$ yields a lower peak frequency compared to the model with the fully consistent temperature description. The simulations with the Shen EoS lead to smaller differences because, as mentioned in Sect.~\ref{sec:sim}, generally lower temperatures are obtained with this EoS. Therefore, thermal effects are not as important as in the case of the LS EoS, and an approximate inclusion has less influence.

Our results are also consistent with the findings summarized in Sect.~\ref{sec:rev}, where we mentioned that the assumption of perfectly efficient cooling of the neutron star merger remnant leads to a more compact remnant structure. For this reason, as a consequence of the missing thermal pressure support, the peak frequencies were found to be higher in \cite{PhysRevD.81.024012}.

Note that the low-frequency domain of the gravitational-wave luminosity spectra is mainly determined by the inspiral phase, where thermal effects are not important. This explains why the spectra in Figs.~\ref{fig:FFT-LS} and ~\ref{fig:FFT-Shen} are very similar at low frequencies.

\subsection{Delay time to black hole collapse}

Besides influencing the oscillation frequency of the postmerger remnant, the different remnant structure also affects the lifetime of the central hypermassive neutron star. Since the remnant rotates differentially, more mass can be stabilized than for a rigid rotator. It therefore collapses to a black hole after angular momentum redistribution. Typically, the delay time $\tau_{\mathrm{delay}}$ between the merging and the collapse is of the order of several 10~ms and depends strongly on the total mass of the system and the EoS \cite{2009arXiv0912.3529D,PhysRevD.81.024012,rezzolla}.

The delay time $\tau_{\mathrm{delay}}$ is an important observational property, because it determines
the maximal length of the gravitational-wave emission by the stable postmerger
remnant. The signal carries information on the EoS of high-density
matter. The gravitational waves produced by the relic black hole
and the surrounding torus are of fundamentally different character,
e.g. at much higher frequencies \cite{2006PhRvD..73f4027S,2008PhRvD..78h4033B}. Finally, the delay time itself may be
an interesting measure as it depends sensitively on the binary system
parameters and the EoS. Consequently, an accurate determination of this
property by simulations is required to find out more about the nature
of neutron star mergers and of supranuclear matter. 

The collapse to a black hole takes place after $\tau_{\mathrm{delay}}=10.3$~ms in the case of our model with $\Gamma_{\mathrm{th}}=1.5$ and the cold LS EoS (see Tab.~\ref{tab:models}). Remarkably, in the simulation using the full temperature dependence a black hole is formed only after 20.8~ms. The difference is a consequence of the higher compactness of the central object in the $\Gamma_{\mathrm{th}}=1.5$-run. Simulating thermal effects with an ideal-gas index of $\Gamma_{\mathrm{th}}=2$ leads to a delay time of 21.0~ms, which is in good agreement with the calculation employing the full temperature-dependent EoS. We assume that this consistency is a consequence of the fact that the structure of the inner region of the remnant, where the collapse sets in, is better reproduced by the $\Gamma_{\mathrm{th}}=2$ choice (see Fig.~\ref{fig:encgth}). The uncertainties of $\tau_{\mathrm{delay}}$ are important findings because they mean that, depending on the specific choice of $\Gamma_{\mathrm{th}}$, the delay time as a very basic property of the merger remnant may be incorrect by a factor of about 2 in simulations using an ideal-gas ansatz.

Because of the high computational costs of our long-term calculations, we cannot determine the delay times for our models based on the Shen EoS. The merger remnants in these models do not collapse to black holes during the simulations. This result is clear, because the Shen EoS leads to less compact central objects and it supports higher neutron star masses in comparison to the LS EoS. Both delays the formation of a black hole. In Tab.~\ref{tab:models} we give lower bounds on the lifetime of the hypermassive remnants for the Shen EoS.

\subsection{Torus masses}
Besides gravitational waves, another characteristic feature of neutron
star mergers of observational relevance is the amount of matter
forming a torus around the black hole emerging from the collapsing central
object. These systems might be the sources of short gamma-ray bursts, provided the energy release from the black hole-torus configuration launches a relativistic outflow producing
gamma-rays by internal shocks. In the context of this
scenario the torus mass is a crucial quantity, because it
determines the possible energy release. This torus mass depends on the binary parameters and the EoS~\cite{2006MNRAS.368.1489O,2007A&A...467..395O,PhysRevD.81.024012}.

The torus mass can be estimated by the requirement that the specific angular momentum of torus matter must be larger than the value at the innermost stable circular orbit, i.e. the torus mass can be roughly determined
before the central object collapses into a black hole. A description of the iterative
procedure is given in~\cite{2007A&A...467..395O}. The formation of the black hole cannot be followed by our code, which is why we monitor the amount of matter fulfilling the torus criterion as a function of time and determine the torus mass shortly before the onset of the gravitational collapse.

The estimated torus masses are listed in Tab.~\ref{tab:models}. The values
increase towards the end of the simulations, because
the merger remnants are not yet stationary. In all six simulations we
observe a continuous contraction of the central object.

Only in the calculations with the LS EoS we can determine the final torus mass, because only in these cases the gravitational collapse of the merger remnant occurs within the simulation time. The torus mass is influenced by two effects, which can be disentangled by monitoring the temporal evolution of the amount of matter fulfilling the torus criterion. After the initial plunge the former cores of the stars bounce and reexpand to contract again several times until these oscillations are damped after some cycles. In the first expansion phase the more compact remnant of the $\Gamma_{\mathrm{th}}=1.5$ model leads to a strong initial increase of the amount of potential torus matter, while this effect is strongly suppressed in the $\Gamma_{\mathrm{th}}=2$ simulation. The fully consistent model with an intermediate compactness shows a moderate growth during the first decompression stage. (We find the same effect comparing the $T=0$ cases and the consistent models discussed in \cite{PhysRevD.81.024012}.) In the subsequent evolution the amount of matter with sufficient angular momentum to fulfill the torus criterion steadily increases with roughly a constant, but lower and similar rate in all models. Therefore, the delay time until black hole formation affects sensitively the final torus mass. Although starting with an initially lower amount of matter obeying the torus criterion, the model with $\Gamma_{\mathrm{th}}=2$ can therefore end up with a higher final value compared to the $\Gamma_{\mathrm{th}}=1.5$ case, because the latter model collapses to a black hole relatively early. The highest final torus mass is obtained for the simulation with the full temperature dependence of the EoS, because the initial amount of torus matter from the first expansion is higher than the one of the $\Gamma_{\mathrm{th}}=2$ calculation, and the delay times are comparable, admitting a long phase of angular momentum transport to the outer parts of the merger remnant.

For the models based on the Shen EoS the amount of matter fulfilling the torus criterion 16~ms after the merging is given in Tab.~\ref{tab:models}. Note that these values are only very crude estimates for the final torus masses, because this quantity depends on the delay time, which cannot be determined in our simulations with the Shen EoS.

For both EoSs the evolution of the torus masses in the simulations with
$\Gamma_{\mathrm{th}}=2$ follows the models with the fully consistent
temperature treatment more closely.

Furthermore, we analyzed the amount of matter that becomes gravitationally unbound from the merger site. This is of interest for the question how neutron star merger events might contribute to the galactic abundances of heavy elements, e.g. by the production of rapid neutron capture nuclei in the ejecta~\cite{2007PhR...450...97A}. We find disagreement in the ejecta masses of about a factor of 2 between our reference models and the calculations with the simplified EoS treatments.

\section{Conclusions} \label{sec:con}
We performed simulations of neutron star mergers with microphysical
EoSs employing an approximate description of thermal effects by an ideal-gas extension of the zero-temperature EoSs on the one hand, and using the full temperature dependence of the EoSs on the other
hand. This allows us to assess the limitations of the approximate
approach and in particular to quantify the deviations occurring in
observational quantities like the gravitational-wave signal. For characterizing the gravitational-wave emission of the postmerger remnant, which is the phase of the event influenced most by
thermal effects, we determine the dominant gravitational-wave frequency of the oscillations of the forming hypermassive object. It is found that the simplified inclusion of thermal effects yields values that differ by about 50 to 250~Hz from the value obtained in the simulation with the fully consistent temperature treatment. The differences can be understood as a consequence of the different remnant compactness, which is caused by the different strengths of the thermal pressure contributions.

Furthermore, the simplified approach affects sensitively the delay time between merging and the formation of a black hole. Again, the discrepancies can be explained by the different remnant structures, where a more compact object collapses to a black hole earlier. The lifetime of the hypermassive remnant can deviate from its true value by a factor of 2 in simulations that use an ideal-gas like ansatz to describe thermal effects. These uncertainties are in particular important, because they influence also the duration of the gravitational-wave emission from the postmerger remnant and thus the integrated power in the relevant frequency bands, which provide information about the high-density EoS. In addition, errors in the delay time also have an effect on the estimates of the amount of matter remaining outside the black hole and the forming accretion torus after the gravitational collapse of the central object.

Our results show that estimates of this disk mass by means of simulations that use the ideal-gas approximation, are uncertain by up to 30\% of the full temperature result. These uncertainties arise because of the effects of the approximate EoS treatment on the delay time as well as the remnant structure and evolution.

The comparison of simulations with the LS EoS suggests that $\Gamma_{\mathrm{th}}=2$ might be a better choice in particular to model the delay time, although an agreement in all relevant quantities cannot be achieved. For instance, considering the frequency of the postmerger ringdown, $\Gamma_{\mathrm{th}}=1.75$ may yield more accurate results. However, conclusions based on such a small number of models (with respect to the investigated EoSs and the binary configurations) have only a limited significance, for which reason we refrain from specifying optimal values for different needs here.

In summary, we stress that the interpretation of results from simulations using the approximate implementation of thermal pressure contributions should take into account the uncertainties presented in this study (see Tab.~\ref{tab:models}). This has an impact in particular on attempts to constrain the high-density EoS by observational data from merging events.

In addition, we expect that for more massive binary configurations the deviations due to the simplified treatment even increase, because mergers of such systems yield higher temperatures \cite{2007A&A...467..395O,PhysRevD.81.024012}.

We note that our findings apply also to other than the investigated zero-temperature EoSs when an ideal-gas ansatz is used to account for temperature effects. For instance, it was shown that piecewise polytropic EoSs can be chosen in order to match microphysical zero temperature EoSs \cite{2009PhRvD..79l4032R}. Also in this case the thermal behavior of the EoS needs to be implemented in an approximate manner giving rise to the associated uncertainties. Therefore, we conclude that microphysical EoSs that include thermal effects consistently are highly demanded for astrophysical applications as the ones discussed here, in particular if a high quantitative accuracy is required.

\begin{acknowledgments}
This work was supported by the Sonderforschungsbereich Transregio 7 ``Gravitational Wave Astronomy", by the Sonderforschungsbereich Transregio 27 ``Neutrinos and Beyond", and the Cluster of Excellence EXC 153 ``Origin and Structure of the Universe" of the Deutsche Forschungsgemeinschaft, and by CompStar, a research networking programme of the European Science Foundation. The computations were performed at the Rechenzentrum Garching of the Max-Planck-Gesellschaft and at the Leibniz-Rechenzentrum Garching.
\end{acknowledgments}

\bibliography{references}

\begin{thebibliography}{32}
\expandafter\ifx\csname natexlab\endcsname\relax\def\natexlab#1{#1}\fi
\expandafter\ifx\csname bibnamefont\endcsname\relax
  \def\bibnamefont#1{#1}\fi
\expandafter\ifx\csname bibfnamefont\endcsname\relax
  \def\bibfnamefont#1{#1}\fi
\expandafter\ifx\csname citenamefont\endcsname\relax
  \def\citenamefont#1{#1}\fi
\expandafter\ifx\csname url\endcsname\relax
  \def\url#1{\texttt{#1}}\fi
\expandafter\ifx\csname urlprefix\endcsname\relax\def\urlprefix{URL }\fi
\providecommand{\bibinfo}[2]{#2}
\providecommand{\eprint}[2][]{\url{#2}}

\bibitem[{\citenamefont{{Duez}}(2010)}]{2009arXiv0912.3529D}
\bibinfo{author}{\bibfnamefont{M.~D.} \bibnamefont{{Duez}}},
  \bibinfo{journal}{Classical and Quantum Gravity}
  \textbf{\bibinfo{volume}{27}}, \bibinfo{pages}{114002}
  (\bibinfo{year}{2010}), \eprint{0912.3529}.

\bibitem[{\citenamefont{{Stairs}}(2004)}]{2004Sci...304..547S}
\bibinfo{author}{\bibfnamefont{I.~H.} \bibnamefont{{Stairs}}},
  \bibinfo{journal}{Science} \textbf{\bibinfo{volume}{304}},
  \bibinfo{pages}{547} (\bibinfo{year}{2004}).

\bibitem[{\citenamefont{{Kalogera} et~al.}(2004)\citenamefont{{Kalogera},
  {Kim}, {Lorimer}, {Burgay}, {D'Amico}, {Possenti}, {Manchester}, {Lyne},
  {Joshi}, {McLaughlin} et~al.}}]{2004ApJ...601L.179K}
\bibinfo{author}{\bibfnamefont{V.}~\bibnamefont{{Kalogera}}},
  \bibinfo{author}{\bibfnamefont{C.}~\bibnamefont{{Kim}}},
  \bibinfo{author}{\bibfnamefont{D.~R.} \bibnamefont{{Lorimer}}},
  \bibinfo{author}{\bibfnamefont{M.}~\bibnamefont{{Burgay}}},
  \bibinfo{author}{\bibfnamefont{N.}~\bibnamefont{{D'Amico}}},
  \bibinfo{author}{\bibfnamefont{A.}~\bibnamefont{{Possenti}}},
  \bibinfo{author}{\bibfnamefont{R.~N.} \bibnamefont{{Manchester}}},
  \bibinfo{author}{\bibfnamefont{A.~G.} \bibnamefont{{Lyne}}},
  \bibinfo{author}{\bibfnamefont{B.~C.} \bibnamefont{{Joshi}}},
  \bibinfo{author}{\bibfnamefont{M.~A.} \bibnamefont{{McLaughlin}}},
  \bibnamefont{et~al.}, \bibinfo{journal}{Astrophys. J. Lett.}
  \textbf{\bibinfo{volume}{601}}, \bibinfo{pages}{L179} (\bibinfo{year}{2004}).

\bibitem[{\citenamefont{{Kim} et~al.}(2006)\citenamefont{{Kim}, {Kalogera}, and
  {Lorimer}}}]{2006astro.ph..8280K}
\bibinfo{author}{\bibfnamefont{C.}~\bibnamefont{{Kim}}},
  \bibinfo{author}{\bibfnamefont{V.}~\bibnamefont{{Kalogera}}},
  \bibnamefont{and} \bibinfo{author}{\bibfnamefont{D.~R.}
  \bibnamefont{{Lorimer}}}, \bibinfo{journal}{ArXiv e-prints, astro-ph/0608280}
   (\bibinfo{year}{2006}), \eprint{arXiv:astro-ph/0608280}.

\bibitem[{\citenamefont{{Oechslin} et~al.}(2007)\citenamefont{{Oechslin},
  {Janka}, and {Marek}}}]{2007A&A...467..395O}
\bibinfo{author}{\bibfnamefont{R.}~\bibnamefont{{Oechslin}}},
  \bibinfo{author}{\bibfnamefont{H.-T.} \bibnamefont{{Janka}}},
  \bibnamefont{and} \bibinfo{author}{\bibfnamefont{A.}~\bibnamefont{{Marek}}},
  \bibinfo{journal}{Astron. Astrophys.} \textbf{\bibinfo{volume}{467}},
  \bibinfo{pages}{395} (\bibinfo{year}{2007}).

\bibitem[{\citenamefont{{Baiotti} et~al.}(2008)\citenamefont{{Baiotti},
  {Giacomazzo}, and {Rezzolla}}}]{2008PhRvD..78h4033B}
\bibinfo{author}{\bibfnamefont{L.}~\bibnamefont{{Baiotti}}},
  \bibinfo{author}{\bibfnamefont{B.}~\bibnamefont{{Giacomazzo}}},
  \bibnamefont{and}
  \bibinfo{author}{\bibfnamefont{L.}~\bibnamefont{{Rezzolla}}},
  \bibinfo{journal}{\prd} \textbf{\bibinfo{volume}{78}},
  \bibinfo{pages}{084033} (\bibinfo{year}{2008}), \eprint{0804.0594}.

\bibitem[{\citenamefont{Bauswein et~al.}(2010)\citenamefont{Bauswein, Oechslin,
  and Janka}}]{PhysRevD.81.024012}
\bibinfo{author}{\bibfnamefont{A.}~\bibnamefont{Bauswein}},
  \bibinfo{author}{\bibfnamefont{R.}~\bibnamefont{Oechslin}}, \bibnamefont{and}
  \bibinfo{author}{\bibfnamefont{H.-T.} \bibnamefont{Janka}},
  \bibinfo{journal}{Phys. Rev. D} \textbf{\bibinfo{volume}{81}},
  \bibinfo{pages}{024012} (\bibinfo{year}{2010}).

\bibitem[{\citenamefont{{Haensel} et~al.}(2007)\citenamefont{{Haensel},
  {Potekhin}, and {Yakovlev}}}]{2007ASSL..326.....H}
\bibinfo{author}{\bibfnamefont{P.}~\bibnamefont{{Haensel}}},
  \bibinfo{author}{\bibfnamefont{A.~Y.} \bibnamefont{{Potekhin}}},
  \bibnamefont{and} \bibinfo{author}{\bibfnamefont{D.~G.}
  \bibnamefont{{Yakovlev}}}, \emph{\bibinfo{title}{{Neutron Stars 1}}}
  (\bibinfo{publisher}{Springer-Verlag, New York}, \bibinfo{year}{2007}).

\bibitem[{\citenamefont{Lattimer and Prakash}(2004)}]{Lattimer:2004pg}
\bibinfo{author}{\bibfnamefont{J.~M.} \bibnamefont{Lattimer}} \bibnamefont{and}
  \bibinfo{author}{\bibfnamefont{M.}~\bibnamefont{Prakash}},
  \bibinfo{journal}{Science} \textbf{\bibinfo{volume}{304}},
  \bibinfo{pages}{536} (\bibinfo{year}{2004}).

\bibitem[{\citenamefont{{Lattimer} and {Prakash}}(2007)}]{2007PhR...442..109L}
\bibinfo{author}{\bibfnamefont{J.~M.} \bibnamefont{{Lattimer}}}
  \bibnamefont{and}
  \bibinfo{author}{\bibfnamefont{M.}~\bibnamefont{{Prakash}}},
  \bibinfo{journal}{Phys. Rep.} \textbf{\bibinfo{volume}{442}},
  \bibinfo{pages}{109} (\bibinfo{year}{2007}), \eprint{arXiv:astro-ph/0612440}.

\bibitem[{\citenamefont{{Steiner} et~al.}(2010)\citenamefont{{Steiner},
  {Lattimer}, and {Brown}}}]{2010arXiv1005.0811S}
\bibinfo{author}{\bibfnamefont{A.~W.} \bibnamefont{{Steiner}}},
  \bibinfo{author}{\bibfnamefont{J.~M.} \bibnamefont{{Lattimer}}},
  \bibnamefont{and} \bibinfo{author}{\bibfnamefont{E.~F.}
  \bibnamefont{{Brown}}}, \bibinfo{journal}{ArXiv e-prints}
  (\bibinfo{year}{2010}), \eprint{1005.0811}.

\bibitem[{\citenamefont{{Lattimer} and {Douglas
  Swesty}}(1991)}]{1991NuPhA.535..331L}
\bibinfo{author}{\bibfnamefont{J.~M.} \bibnamefont{{Lattimer}}}
  \bibnamefont{and} \bibinfo{author}{\bibfnamefont{F.}~\bibnamefont{{Douglas
  Swesty}}}, \bibinfo{journal}{Nuclear Physics A}
  \textbf{\bibinfo{volume}{535}}, \bibinfo{pages}{331} (\bibinfo{year}{1991}).

\bibitem[{\citenamefont{{Shen} et~al.}(1998)\citenamefont{{Shen}, {Toki},
  {Oyamatsu}, and {Sumiyoshi}}}]{1998NuPhA.637..435S}
\bibinfo{author}{\bibfnamefont{H.}~\bibnamefont{{Shen}}},
  \bibinfo{author}{\bibfnamefont{H.}~\bibnamefont{{Toki}}},
  \bibinfo{author}{\bibfnamefont{K.}~\bibnamefont{{Oyamatsu}}},
  \bibnamefont{and}
  \bibinfo{author}{\bibfnamefont{K.}~\bibnamefont{{Sumiyoshi}}},
  \bibinfo{journal}{Nuclear Physics A} \textbf{\bibinfo{volume}{637}},
  \bibinfo{pages}{435} (\bibinfo{year}{1998}), \eprint{arXiv:nucl-th/9805035}.

\bibitem[{\citenamefont{{Shibata}}(2005)}]{2005PhRvL..94t1101S}
\bibinfo{author}{\bibfnamefont{M.}~\bibnamefont{{Shibata}}},
  \bibinfo{journal}{Phys. Rev. Lett.} \textbf{\bibinfo{volume}{94}},
  \bibinfo{pages}{201101} (\bibinfo{year}{2005}), \eprint{arXiv:gr-qc/0504082}.

\bibitem[{\citenamefont{{Shibata} et~al.}(2005)\citenamefont{{Shibata},
  {Taniguchi}, and {Ury{\= u}}}}]{2005PhRvD..71h4021S}
\bibinfo{author}{\bibfnamefont{M.}~\bibnamefont{{Shibata}}},
  \bibinfo{author}{\bibfnamefont{K.}~\bibnamefont{{Taniguchi}}},
  \bibnamefont{and} \bibinfo{author}{\bibfnamefont{K.}~\bibnamefont{{Ury{\=
  u}}}}, \bibinfo{journal}{\prd} \textbf{\bibinfo{volume}{71}},
  \bibinfo{pages}{084021} (\bibinfo{year}{2005}), \eprint{arXiv:gr-qc/0503119}.

\bibitem[{\citenamefont{{Shibata} and {Taniguchi}}(2006)}]{2006PhRvD..73f4027S}
\bibinfo{author}{\bibfnamefont{M.}~\bibnamefont{{Shibata}}} \bibnamefont{and}
  \bibinfo{author}{\bibfnamefont{K.}~\bibnamefont{{Taniguchi}}},
  \bibinfo{journal}{\prd} \textbf{\bibinfo{volume}{73}},
  \bibinfo{pages}{064027} (\bibinfo{year}{2006}).

\bibitem[{\citenamefont{{Kiuchi} et~al.}(2009)\citenamefont{{Kiuchi},
  {Sekiguchi}, {Shibata}, and {Taniguchi}}}]{2009PhRvD..80f4037K}
\bibinfo{author}{\bibfnamefont{K.}~\bibnamefont{{Kiuchi}}},
  \bibinfo{author}{\bibfnamefont{Y.}~\bibnamefont{{Sekiguchi}}},
  \bibinfo{author}{\bibfnamefont{M.}~\bibnamefont{{Shibata}}},
  \bibnamefont{and}
  \bibinfo{author}{\bibfnamefont{K.}~\bibnamefont{{Taniguchi}}},
  \bibinfo{journal}{\prd} \textbf{\bibinfo{volume}{80}},
  \bibinfo{pages}{064037} (\bibinfo{year}{2009}), \eprint{0904.4551}.

\bibitem[{\citenamefont{Abbott et~al.}(2009)}]{Abbott:2007kv2}
\bibinfo{author}{\bibfnamefont{B.}~\bibnamefont{Abbott}} \bibnamefont{et~al.}
  (\bibinfo{collaboration}{LIGO Scientific}), \bibinfo{journal}{Rept. Prog.
  Phys.} \textbf{\bibinfo{volume}{72}}, \bibinfo{pages}{076901}
  (\bibinfo{year}{2009}), \eprint{0711.3041}.

\bibitem[{\citenamefont{Acernese et~al.}(2006)}]{Acernese:2006bj}
\bibinfo{author}{\bibfnamefont{F.}~\bibnamefont{Acernese}}
  \bibnamefont{et~al.}, \bibinfo{journal}{Class. Quant. Grav.}
  \textbf{\bibinfo{volume}{23}}, \bibinfo{pages}{S635} (\bibinfo{year}{2006}).

\bibitem[{\citenamefont{{Oechslin} and {Janka}}(2006)}]{2006MNRAS.368.1489O}
\bibinfo{author}{\bibfnamefont{R.}~\bibnamefont{{Oechslin}}} \bibnamefont{and}
  \bibinfo{author}{\bibfnamefont{H.-T.} \bibnamefont{{Janka}}},
  \bibinfo{journal}{Mon. Not. R. Astron. Soc.} \textbf{\bibinfo{volume}{368}},
  \bibinfo{pages}{1489} (\bibinfo{year}{2006}).

\bibitem[{\citenamefont{{Oechslin} and {Janka}}(2007)}]{2007PhRvL..99l1102O}
\bibinfo{author}{\bibfnamefont{R.}~\bibnamefont{{Oechslin}}} \bibnamefont{and}
  \bibinfo{author}{\bibfnamefont{H.-T.} \bibnamefont{{Janka}}},
  \bibinfo{journal}{Phys. Rev. Lett.} \textbf{\bibinfo{volume}{99}},
  \bibinfo{pages}{121102} (\bibinfo{year}{2007}).

\bibitem[{\citenamefont{{Janka} et~al.}(1993)\citenamefont{{Janka}, {Zwerger},
  and {M\"onchmeyer}}}]{1993A&A...268..360J}
\bibinfo{author}{\bibfnamefont{H.-T.} \bibnamefont{{Janka}}},
  \bibinfo{author}{\bibfnamefont{T.}~\bibnamefont{{Zwerger}}},
  \bibnamefont{and}
  \bibinfo{author}{\bibfnamefont{R.}~\bibnamefont{{M\"onchmeyer}}},
  \bibinfo{journal}{Astron. Astrophys.} \textbf{\bibinfo{volume}{268}},
  \bibinfo{pages}{360} (\bibinfo{year}{1993}).

\bibitem[{\citenamefont{{Dimmelmeier} et~al.}(2005)\citenamefont{{Dimmelmeier},
  {Novak}, {Font}, {Ib{\'a}{\~n}ez}, and {M{\"u}ller}}}]{2005PhRvD..71f4023D}
\bibinfo{author}{\bibfnamefont{H.}~\bibnamefont{{Dimmelmeier}}},
  \bibinfo{author}{\bibfnamefont{J.}~\bibnamefont{{Novak}}},
  \bibinfo{author}{\bibfnamefont{J.~A.} \bibnamefont{{Font}}},
  \bibinfo{author}{\bibfnamefont{J.~M.} \bibnamefont{{Ib{\'a}{\~n}ez}}},
  \bibnamefont{and}
  \bibinfo{author}{\bibfnamefont{E.}~\bibnamefont{{M{\"u}ller}}},
  \bibinfo{journal}{\prd} \textbf{\bibinfo{volume}{71}},
  \bibinfo{pages}{064023} (\bibinfo{year}{2005}),
  \eprint{arXiv:astro-ph/0407174}.

\bibitem[{\citenamefont{{Isenberg} and {Nester}}(1980)}]{1980grg..conf...23I}
\bibinfo{author}{\bibfnamefont{J.}~\bibnamefont{{Isenberg}}} \bibnamefont{and}
  \bibinfo{author}{\bibfnamefont{J.}~\bibnamefont{{Nester}}}, in
  \emph{\bibinfo{booktitle}{General Relativity and Gravitation}}
  (\bibinfo{publisher}{Plenum Press, New York}, \bibinfo{year}{1980}),
  p.~\bibinfo{pages}{23}.

\bibitem[{\citenamefont{{Oechslin} et~al.}(2002)\citenamefont{{Oechslin},
  {Rosswog}, and {Thielemann}}}]{2002PhRvD..65j3005O}
\bibinfo{author}{\bibfnamefont{R.}~\bibnamefont{{Oechslin}}},
  \bibinfo{author}{\bibfnamefont{S.}~\bibnamefont{{Rosswog}}},
  \bibnamefont{and} \bibinfo{author}{\bibfnamefont{F.-K.}
  \bibnamefont{{Thielemann}}}, \bibinfo{journal}{\prd}
  \textbf{\bibinfo{volume}{65}}, \bibinfo{pages}{103005}
  (\bibinfo{year}{2002}).

\bibitem[{\citenamefont{{Belczynski} et~al.}(2008)\citenamefont{{Belczynski},
  {O'Shaughnessy}, {Kalogera}, {Rasio}, {Taam}, and
  {Bulik}}}]{2008ApJ...680L.129B}
\bibinfo{author}{\bibfnamefont{K.}~\bibnamefont{{Belczynski}}},
  \bibinfo{author}{\bibfnamefont{R.}~\bibnamefont{{O'Shaughnessy}}},
  \bibinfo{author}{\bibfnamefont{V.}~\bibnamefont{{Kalogera}}},
  \bibinfo{author}{\bibfnamefont{F.}~\bibnamefont{{Rasio}}},
  \bibinfo{author}{\bibfnamefont{R.~E.} \bibnamefont{{Taam}}},
  \bibnamefont{and} \bibinfo{author}{\bibfnamefont{T.}~\bibnamefont{{Bulik}}},
  \bibinfo{journal}{Astrophys. J. Lett.} \textbf{\bibinfo{volume}{680}},
  \bibinfo{pages}{L129} (\bibinfo{year}{2008}).

\bibitem[{\citenamefont{{Rezzolla} et~al.}(2009)}]{rezzolla}
\bibinfo{author}{\bibfnamefont{L.}~\bibnamefont{{Rezzolla}}}
  \bibnamefont{et~al.}, \emph{\bibinfo{title}{in preparation}}
  (\bibinfo{year}{2009}).

\bibitem[{\citenamefont{{Price}}(2007)}]{2007PASA...24..159P}
\bibinfo{author}{\bibfnamefont{D.~J.} \bibnamefont{{Price}}},
  \bibinfo{journal}{Publications of the Astronomical Society of Australia}
  \textbf{\bibinfo{volume}{24}}, \bibinfo{pages}{159} (\bibinfo{year}{2007}),
  \eprint{0709.0832}.

\bibitem[{\citenamefont{{Blanchet} et~al.}(1990)\citenamefont{{Blanchet},
  {Damour}, and {Schaefer}}}]{1990MNRAS.242..289B}
\bibinfo{author}{\bibfnamefont{L.}~\bibnamefont{{Blanchet}}},
  \bibinfo{author}{\bibfnamefont{T.}~\bibnamefont{{Damour}}}, \bibnamefont{and}
  \bibinfo{author}{\bibfnamefont{G.}~\bibnamefont{{Schaefer}}},
  \bibinfo{journal}{Mon. Not. R. Astron. Soc.} \textbf{\bibinfo{volume}{242}},
  \bibinfo{pages}{289} (\bibinfo{year}{1990}).

\bibitem[{\citenamefont{{Arnould} et~al.}(2007)\citenamefont{{Arnould},
  {Goriely}, and {Takahashi}}}]{2007PhR...450...97A}
\bibinfo{author}{\bibfnamefont{M.}~\bibnamefont{{Arnould}}},
  \bibinfo{author}{\bibfnamefont{S.}~\bibnamefont{{Goriely}}},
  \bibnamefont{and}
  \bibinfo{author}{\bibfnamefont{K.}~\bibnamefont{{Takahashi}}},
  \bibinfo{journal}{Phys. Rep.} \textbf{\bibinfo{volume}{450}},
  \bibinfo{pages}{97} (\bibinfo{year}{2007}), \eprint{0705.4512}.

\bibitem[{\citenamefont{{Read} et~al.}(2009)\citenamefont{{Read}, {Lackey},
  {Owen}, and {Friedman}}}]{2009PhRvD..79l4032R}
\bibinfo{author}{\bibfnamefont{J.~S.} \bibnamefont{{Read}}},
  \bibinfo{author}{\bibfnamefont{B.~D.} \bibnamefont{{Lackey}}},
  \bibinfo{author}{\bibfnamefont{B.~J.} \bibnamefont{{Owen}}},
  \bibnamefont{and} \bibinfo{author}{\bibfnamefont{J.~L.}
  \bibnamefont{{Friedman}}}, \bibinfo{journal}{\prd}
  \textbf{\bibinfo{volume}{79}}, \bibinfo{pages}{124032}
  (\bibinfo{year}{2009}), \eprint{0812.2163}.

\bibitem[{\citenamefont{{Meltzer} and {Thorne}}(1966)}]{1966ApJ...145..514M}
\bibinfo{author}{\bibfnamefont{D.~W.} \bibnamefont{{Meltzer}}}
  \bibnamefont{and} \bibinfo{author}{\bibfnamefont{K.~S.}
  \bibnamefont{{Thorne}}}, \bibinfo{journal}{\apj}
  \textbf{\bibinfo{volume}{145}}, \bibinfo{pages}{514} (\bibinfo{year}{1966}).

\end{thebibliography}

\end{document}